# A Reinforcement Learning Framework with Region-Awareness and Shared Path Experience for Efficient Routing in Networks-on-Chip

Kamil Khan
Electrical and Computer Engineering
Colorado State University
Fort Collins, CO, USA
kamil@colostate.edu

Sudeep Pasricha
Electrical and Computer Engineering
Colorado State University
Fort Collins, CO, USA
sudeep@colostate.edu

*Abstract*—Network-on-chip (NoC) architectures provide a scalable, high-performance, and reliable interconnect for emerging manycore systems. The routing policies used in NoCs have a significant impact on overall performance. Prior efforts have proposed reinforcement learning (RL)-based adaptive routing policies to avoid congestion and minimize latency in NoCs. The output quality of RL policies depends on selecting a representative cost function and an effective update mechanism. Unfortunately, existing RL policies for NoC routing fail to represent path contention and regional congestion in the cost function. Moreover, the experience of packet flows sharing the same route is not fully incorporated into the RL update mechanism. In this paper, we present a novel regional congestion-aware RL-based NoC routing policy called Q-RASP that is capable of sharing experience from packets using the same routes. Q-RASP improves average packet latency by up to 18.3% and reduces NoC energy consumption by up to 6.7% with minimal area overheads compared to state-of-the-art RL-based NoC routing implementations.

## I. Introduction

As manycore chips continue to integrate an increasing number of cores, network-on-chip (NoC) architectures have emerged as the leading solution to overcome data movement bottlenecks by offering improved scalability, performance, and reliability during communication [1]. Within NoCs, routing policies play a key role in determining the path that packets take towards a destination. Traditional routing policies range from simple deterministic policies such as dimension-order XY routing which always chooses the same route for packets between a source and destination, to more complex adaptive congestion-aware reinforcement learning (RL) based policies such as Q-routing [2].

Typically, routing policies can be classified as oblivious or adaptive based on how routing decisions are made. Oblivious routing policies do not consider the current network state when selecting routing paths, e.g., XY routing, random walk routing, etc. In contrast, adaptive policies consider runtime knowledge, such as network congestion information based on downstream router buffer occupancy, when selecting a routing path. Typically, adaptive routing can perform better than oblivious routing when packet flows are non-uniform, i.e., when some nodes are over-represented in the communication traffic. In such cases, adaptive routing can better distribute traffic in the NoC to avoid congestion.

While adaptive policies can greedily select the least congested next hop, they lack information about congestion on the remaining path. The problem can be solved by communicating the global state of congestion in the NoC to every router, but such solutions scale poorly and incur excessive communication overhead. Q-routing [3] represents an alternative solution whereby the routing policy can form a global estimate of congestion by using information only from neighboring nodes. A NoC router is associated with each node in the network. The router maintains a table with congestion estimates for different destinations in the form of Q-values. Q-routing policies have been shown to improve NoC performance compared to other routing policies [3] due to their ability to learn from experience.

The performance of Q-routing policies depends on their 1) cost function, and 2) update mechanism. The cost function defines how congestion is inferred from the network condition to assess the quality of a routing action, whereas the update mechanism defines a methodology to use the experience of the routing agent (cost and estimates from neighbor nodes) to update the existing policy. In prior Q-routing approaches, the cost can be slow to calculate, or fail to represent congestion accurately. Further, update mechanisms used in prior approaches fail to effectively update the Q-values, resulting in underperforming Q-routing policies.

We address these limitations by designing an improved Q-learning based NoC routing framework called Q-RASP (Q-Routing for NoCs with Region-Awareness and Shared Path Experience). Q-RASP defines a path- and region-aware cost by prioritizing path contention over latency or input contention and augmenting it with a regional cost component. It also shares learning updates between packet flows to different destinations which share the same route. Our work makes the following novel contributions:

- We identify the shortcomings of existing NoC Q-routing policies in terms of cost functions and update mechanisms.
- We develop a new region-aware path contention-based NoC Q-routing cost function for selecting contention-free paths in congestion-free regions.
- We devise a new update mechanism to share learning experience between packet flows with different destinations using the same routes.

The rest of this paper is organized as follows. Section II presents a background on Q-learning and Q-routing. Section III discusses prior work. Section IV describes our Q-routing framework, Q-RASP. Section V presents experimental results followed by concluding remarks in Section VI.

## II. BACKGROUND

Q-routing is a class of adaptive routing policies which uses the learning framework of Q-learning to solve the routing problem in NoCs as a reinforcement learning (RL) problem. We discuss Q-learning as well as Q-routing for NoCs (which uses Q-learning) in the rest of this section.

### A. Q-learning

Q-learning is a model-free learning algorithm for solving RL problems. It is model-free because it does not require a model of the environment to learn a policy. In RL, an agent learns to maximize the return, or long-term reward, over a sequence of actions in an environment. The agent has a representation of the current environment state, $s$. At each step $t$, the agent takes an action $a_t$ in state $s_t$. The action transitions the agent state from $s_t$ to the next state, $s_{t+1}$. At step $t + 1$, the agent also receives a scalar reward for the action $a_t$ which is used by a learning algorithm like Q-learning to update the current agent policy.

In Q-learning, an agent uses the state-value function $Q(s_t, a_t)$ to estimate the expected return of an action $a_t$ in state $s_t$. As each step $t$, the policy is updated using Eq. (1):

$$Q^{new}(s_t, a_t) \leftarrow (1 - \alpha) \cdot Q(s_t, a_t) + \alpha \cdot \left( r_t + \gamma \cdot \max_{a \in A} Q(s_{t+1}, a) \right) \quad (1)$$

where $\alpha$ is the learning rate, $r_t$ is the reward for action $a_t$ in state $s_t$, $\gamma$ is the discount factor, and $\max_{a \in A} Q(s_{t+1}, a)$ is the expected return for choosing the best action at state $s_{t+1}$ from the set of all actions $A$. The learning rate specifies the weight of new experience compared to the existing estimate. The reward $r_t$ is a scalar value representing the consequence of taking action $a_t$ at $s_t$. $\gamma \in [0,1]$ is the discount factor that determines the importance of future rewards. With a discount factor ($\gamma$) of 0, the agent will consider only current rewards, represented by $r_t$ in Eq. (1). Conversely, as $\gamma$ approaches 1, the agent will consider future rewards equally.

The function $Q(s, a)$ is commonly represented using a table called the Q-table. The Q-table stores the expected return of each state-action pair as Q-values $Q(s_t, a_t)$. The agent policy can simply be a greedy selection function to maximize reward or minimize cost. For example, an agent policy may maximize the cumulative reward by taking action $a_* = \max_{a \in A} Q(s_t, a)$ at each step. In most implementations, an $\varepsilon$-greedy policy is used where an agent can query the table to either select the greedy action $a_*$ to optimize return or randomly select an action with a probability $\varepsilon$ to explore the environment for learning. In this way, the $\varepsilon$-greedy policy can explore alternative solutions by selecting non-optimal actions with some probability. Given sufficient update coverage of all state-action pairs, the policy eventually converges to the optimal state-action policy.

### B. Q-routing in NoCs

Q-routing applies the RL framework to the problem of routing in NoCs using Q-learning. Typically, the objective is to learn a routing policy which can minimize packet latency by adapting to the network condition online.

In Q-routing, the state, $s$ is defined as the location of the packet being routed (i.e., the local router) and the packet's destination. In the selection stage of routing, the direction of the output port selected at each hop is denoted as the action, $a$. After the packet reaches the downstream router, the downstream router transmits a learning packet back to the local router with its own estimate of the remaining cost until the packet reaches its destination. The estimate of the remaining cost and the immediate cost of selecting action $a$ (experience) is used to update the Q-routing policy to reinforce actions which reduce total packet latency from the source to the destination. The Q-values representing the Q-routing policy are updated using Eq (2):

$$Q_x(d, y) \leftarrow (1 - \alpha) \cdot Q_x(d, y) + \alpha \cdot \left( q_x + \gamma \cdot \min_{z \in Z} Q_y(d, z) \right) \quad (2)$$

Eq. (2) represents the case where router $x$ estimates the latency to send a packet to node $d$ using the neighbor $y$ as $Q_x(d, y)$. When router $x$ selects neighbor $y$ as the next hop, node $y$ must send back its own estimate of minimum latency for the packet's remaining journey $\min_{z \in Z} Q_y(d, z)$, where $Z$ is the set of all valid neighbours of $y$. Finally, router $x$ must update the Q-value $Q_x(d, y)$ using the time the packet spends in the router $x$'s local queue, $q_x$ and the estimate from node $y$. The discount rate $\gamma$ is used to adjust the weight of the congestion estimate for the remaining path in the update. Similarly, the learning rate $\alpha$ can be used to assign weightage to new experiences. Note that the cost $q_x$ is a Q-routing design choice and can be redefined. Furthermore, while Eq. (2) defines a single update per routing action, a Q-routing update mechanism can specify multiple updates to avoid stale Q-values.

TABLE I. TABLE OF Q-VALUES FOR DESTINATIONS 0-3 IN ROUTER 8 IN 8×8 NoC MESH NETWORK WITH MINIMAL-ROUTE Q-ROUTING.

| Destination ($d$) | Neighbors ($y$) | | Q-Values ($Q_8(d, y)$) | |
|---|---|---|---|---|
| Node 0 | Node 0 | | $Q_8(0,0)$ | |
| Node 1 | Node 0 | Node 1 | $Q_8(1,0)$ | $Q_8(1,1)$ |
| Node 2 | Node 0 | Node 1 | $Q_8(2,0)$ | $Q_8(2,1)$ |
| Node 3 | Node 0 | Node 1 | $Q_8(3,0)$ | $Q_8(3,1)$ |

The Q-routing policy is stored in a distributed manner in the form of tables inside routers. The table in each router stores Q-values which represent the estimated latency to each destination node for each output port available for selection. Table I shows an example of the table from Node 8's router in an 8×8 mesh NoC. When Node 8 sends a packet to Node 3 through neighbor Node 0, the Q-value for the routing action $Q_8(3,0)$ will be updated using Eq. (2). Upon receiving the packet from Node 8, Node 0 sends back its own estimate for the packet's remaining journey, i.e., $Q_0(3,1)$, assuming it selects its neighbor Node 1 as the next node. Next, the weighted sum of the latency in Node 8's router $q_x$ and the estimate $Q_0(3,1)$ is used to update $Q_8(3,0)$ in the table using Eq. (2). At each hop the router follows the minimum selection rule, i.e., it selects the action with the lowest Q-value to minimize total packet latency.

The tables with Q-values and the minimum selection rule enable selecting the path with the minimum expected total latency to every destination from every node in the network. By selecting the neighbor with the minimum expected latency, a Q-routing policy aims to minimize the total accumulated cost of its routing actions. In this section, the cost $q_x$ has been defined as the queue latency of the packet in the local router. When defined as such, the Q-values in each router converge to the total latency (cost) from the current router to the destination. Since the Q-values determine the routing policy, the choice of cost function is central to the Q-routing policy. The cost function is flexible and can be redefined to even represent goals different than minimizing latency, e.g., minimizing transmission power, controlling router temperature, etc. As packets are routed using Q-routing, the path with the least total cost to each destination is learned using experience from packets going to the same destination.

### III. RELATED WORK

We focus on prior work that aims to minimize packet latency with Q-routing as described in the previous section. In prior work, cost functions include queue latency of the packet in the local router [2], [4], queue latency of the packet in the downstream router [5]–[7], and input buffer utilization of the downstream router [8], [9]. By selecting neighbors with the lowest expected latency or number of occupied input buffers, contention in the path to the destination is minimized, thus reducing latency.

A Q-routing policy must also ensure that the Q-values in the tables are frequently updated to reflect the most current state of the network. In the Q-routing algorithm, the update of Q-values (Eq. (2)) is triggered by the routing of a packet through a router. But if packets are not frequently sent to a destination over some paths, the Q-values for those paths are not updated. Therefore, the routing table represents an outdated state of the network, resulting in suboptimal routing decisions. To address the problem of outdated Q-values, different modifications to the update mechanism have been proposed. In Bi-LCQ [3], data packets are appended with Q-value estimates to update the Q-values of the path in the reverse direction. In CrQ [6], a credence measure associated with Q-values is used to adjust the weight of each update. The value of credence is high for Q-values that have been updated recently. During the update step, the learning rate is increased for incoming Q-values with high credence.

There are several drawbacks with prior Q-routing implementations. For example, when local queue packet latency is used as the cost, it fails to use information from the downstream router for accurate estimates. When downstream packet latency is used, it is only available after the packet exits the downstream router. The delay in cost availability reduces the speed of learning. This delay can be avoided if the buffer status of the downstream router is used instead of queue latency. However, the packet also faces contention for the output port of the downstream router which is not considered in prior work. The update mechanism used by techniques such as Bi-LCQ can additionally update the path in the reverse direction of a packet flow. However, the reverse flow may not exist in all traffic scenarios and in such cases, the Q-values may never be used. Lastly, large variations in the learning rate in techniques such as CrQ can lead to unstable learning and the policy may never converge.

### IV. Q-RASP OVERVIEW

In this section, we describe our NoC Q-routing framework Q-RASP which overcomes the drawbacks of prior work. First, we introduce our new path and region contention-aware cost function. Then we describe our new experience-sharing update mechanism. Lastly, we discuss the algorithm's implementation.

#### A. Path- And Region-aware Cost Function

The definition of cost plays an important role in the Q-routing policy. The choice of using a cost associated with the downstream router instead of the current router reduces the number of estimated hop latencies by one. However, using a delayed cost such as packet latency in the input queue of the downstream router causes a delay in the update after the action has been taken. This is because packet latency is only available after the switch traversal stage in the downstream router. The delayed updates reduce the ability of the policy to adapt quickly to the network condition. To avoid the delay associated with latency, the cost can be derived from the input buffer/virtual channel (VC) utilization of the downstream router immediately after the packet enters the neighboring router. The input buffer utilization represents contention at the input port of the downstream router.

In addition to contention at the input port, the packet's latency is also affected by contention at the output port. Information about the output port contention such as the number of reserved VCs or the number of occupied buffers is available from the VC reservation table or the credits table, respectively, in the neighboring router. By augmenting the input port contention with output port contention, we define a path contention cost, $q_p$ as:

$$q_p = r_i + r_o \quad (3)$$

where $r_i$ is the number of occupied VCs at the input, and $r_o$ is the number of reserved VCs at the output port.

We also introduce an additional cost of region contention. The introduction of the region congestion cost, $q_r$ allows the Q-values to be influenced by the cost for alternative paths which are not currently in use, i.e., path contention for paths in the region other than the currently chosen path. A regional component in the cost reinforces the policy to find contention-free paths passing through less congested regions. The ability to select paths which pass through less congested regions enables Q-RASP to quickly shift to a nearby optimal route when some part of the path becomes congested. We define the region contention cost $q_r$ as the sum of contention for all possible output ports:

$$q_r = \sum_O r_o \quad (4)$$

where $q_r$ is region contention cost, $O$ is the set of all routing options, and $r_o$ is the number of reserved VCs in the output direction $o$. Lastly, the total cost is calculated using a weighted sum of the path and region contention cost components. We use the Q-routing update from Eq. (2) and replace the cost $q_x$ with $q_y$:

$$q_y = q_p + \mu q_r \quad (5)$$

where $\mu \in [0, 1]$ and is used for assigning priority to the region cost component.

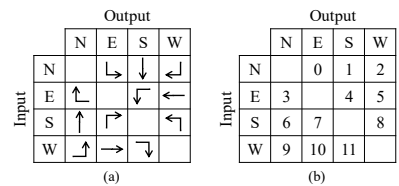

Figure 1. All possible routes through a router in a 2D-mesh NoC shown as (a) route arrows, (b) route number values.

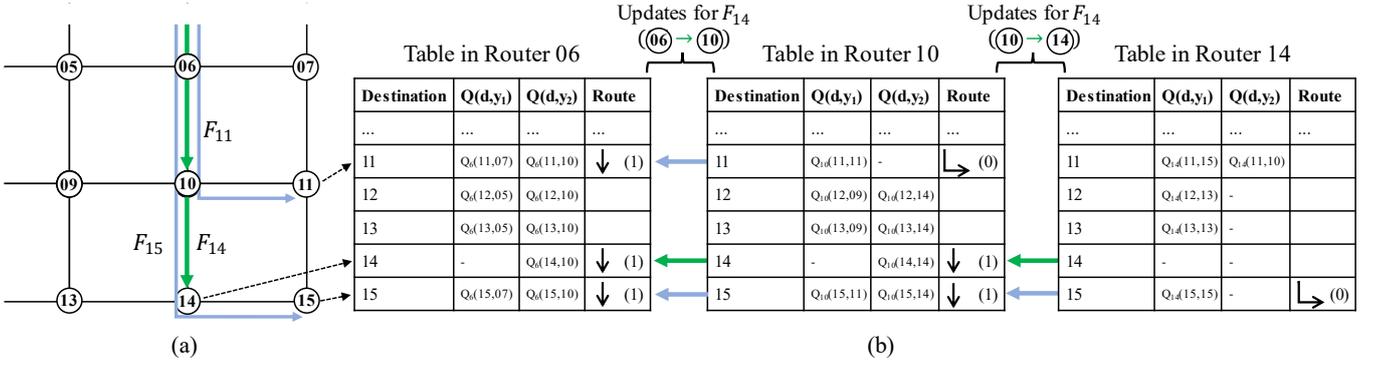

Figure 2. (a) Packet flows $F_{11}$, $F_{15}$, and $F_{14}$ sharing routes while going to different destinations. (b) Q-RASP update mechanism using the experience from routing flow $F_{14}$ to update the Q-values associated with packet flows $F_{11}$, and $F_{15}$.

## B. Shared Path Experience Update Mechanism

As discussed in Section II, for Q-routing to consistently find optimal routes, the Q-values must frequently be updated to represent the current state of congestion in the network. In Q-routing, if packets to a destination $d$ in router $x$ have not used neighbor $y$ recently, $Q_x(d,y)$ will not be updated and continue to represent a prior state of the network. However, we recognize that $Q_x(d,y)$ can be updated using the cost experienced by packets going to destinations other than $d$ which use the same route through the router, where a route is defined by the input and output ports used by a packet (See Fig. 1).

As an example, Fig. 2(a) shows a section of a 4×4 2D-mesh NoC with traffic flows $F_{11}$, $F_{15}$, and $F_{14}$. Here $F_{11}$ and $F_{15}$ have been routed before $F_{14}$. Traffic flow $F_{11}$ has been routed to destination node 11, $F_{15}$ to destination node 15, and $F_{14}$ is currently being routed to destination node 14. We follow $F_{14}$ as a packet is routed first from node 06 to node 10, and then from node 10 to node 14. Here, $F_{14}$ shares the first hop's route (↓) with both $F_{11}$ and $F_{15}$, and the second hop's route (↓) with $F_{15}$. Fig. 2(b) shows the tables of Q-values from router 06, 10, and 14. For each destination, the table lists two Q-values, one for neighbor $y_1$ and another for neighbor $y_2$. Note that as we consider the minimal routing path to the destination, each node can have a maximum of two possible neighbors to choose from. Each table of Q-values must therefore have two columns of Q-values, one for each possible neighbor. We represent the updates made to each table as $F_{14}$ travels along the path 06 → 10 → 14 with arrows between the table rows. While the updates to the Q-values for node 14 are common in all Q-routing policies, we propose Shared Path Experience for additional updates for nodes 11 and 15 due to the shared paths between the flows. Our update mechanism ensures that when packets belonging to flow $F_{11}$ and $F_{15}$ are routed in the future, the corresponding Q-values represent a more current state of the network congestion.

In order to use the additional updates, we extend the table in each router by a column to remember the routes taken by packet flows to each destination. The route through a router is denoted by the pair (Input Port, Output Port), and represented using integer values as shown in Fig. 1(b). In summary, when a packet destined to (destination) node $d$ is sent from router $x$ to neighbor node $y$, node $y$ must send back the estimate $Q_y(d,z)$ for destination $d$, as well as $Q_y(d',z)$ for each destination $d'$ which has the same value in the router table's "Route" column as the current packet. Upon receiving the learning packet, router $x$ must use the cost $q_y$ representing congestion at router $y$ and the Q-value estimates for all destinations to update the corresponding Q-values in its table using Eq. (2).

## C. Implementation

We implement the proposed routing algorithm by adding logic for calculating the new cost and storing additional information in the table for the Shared Path Experience update mechanism. As discussed earlier, the total cost for a routing action is the combination of the path contention cost and the region contention cost (see Eq. (3)). We use comparators, adders, and a multiplier alongside the existing input VCs and the reservation table to calculate the total cost. For the Shared Path Experience update mechanism, we must store route information for each packet flow to identify flows that share the same route through the router.

To calculate the path contention cost, we use two sets of comparators: one to check if input VCs are occupied and another to check the reservation status of the output VCs. The outputs of the comparators increment the partial sums $r_i$, the total number of occupied input VCs and $r_o$, the total number of reserved output VCs. Next, we use a single 4-bit adder to output the total path contention cost $q_p$ as the sum of $r_i$ and $r_o$. The second component of the cost, region contention cost $q_r$, is found by adding the number of reserved VCs in the outputs other than the selected output $o$ to $q_p$. The path contention cost and region contention cost are combined into a weighted sum $q_y$ by a 4×4 multiplier using the weight $\mu$ (see Eq. (5)). Lastly, the cost $q_y$ is written into a learning packet with the estimate $\min_{z \in Z} Q_y(d,z)$ to be used in the update. For transmitting the learning packets, we use dedicated links to avoid contention with data packets. The process of routing, packet traversal, cost calculation, and update is shown in steps 1-4 in Fig. 3.

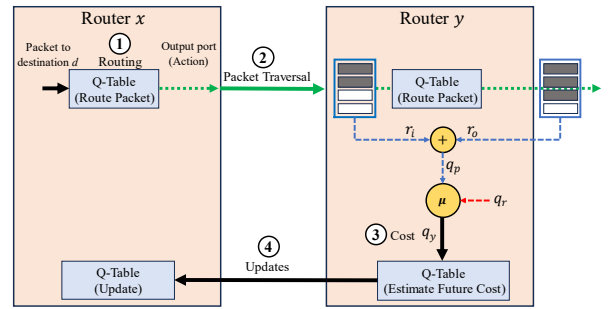

Figure 3. Steps 1-4 show the routing and policy update sequence for Q-RASP. Router $x$ routes a packet to neighbor node $y$, and node $y$ returns one or more learning packets for updates.

For implementing the modified table with the proposed update mechanism, we extend the tables by one column for storing the route taken by each packet flow. The total number of possible routes is determined by the number of input and output channels for each packet flow. For example, when using minimal-route Q-routing, the route for any destination has a maximum of two input and two output options, which results in 4 total route options. Therefore, an additional 3 bits of information are added to each row for tracking routes for each destination. On the other hand, our choice of cost function reduces the total number of bits required for the fixed-point representation of Q-values. For example, if queue latency is used as the cost, the Q-values represent the expected latency of the remaining journey of the packet. The upper limit for total latency can be very large under high traffic loads, requiring a greater number of integer bits to represent the whole range. In our implementation, the Q-values represent the total number of packets competing for the remaining path in the region. The upper limit in this case is a linear function of the number of nodes and channels in the network and is much smaller in most cases than the upper bound of latency. Therefore, we use a fixed-point representation with just 6 bits for the integer part, and 4 bits for the fractional part.

Once the additional updates have been identified using the route information and written into learning packets, we use a buffer queue to store them. For our experiments, we chose a queue size of 4 flits to buffer 4 single-flit learning packets. The buffer size can be adjusted to trade-off storage/transmission overheads and the rate of update.

We use partially-adaptive minimal-route Q-routing to avoid livelock and reduce the overhead of the table for storing Q-values by reducing the number of routing actions. For deadlock freedom, partially-adaptive algorithms employ turn-models [10], [11] to restrict some turns. The purpose of restricting some turns is to avoid packets waiting for each other in a cycle which causes deadlock. In our implementation, we split the available VCs into two sets, each with different turns restricted. In one set of VCs, south-first turns are restricted, whereas in the other set, north-first turns are restricted.

## V. EXPERIMENTAL RESULTS

### A. Experimental Setup

We use the NoC simulator Noxim [12] to evaluate the performance of the proposed Q-RASP framework. We implemented the table and combinatorial logic for Q-learning in RTL and synthesized it using Cadence Genus to obtain power and area estimates. For estimating the area of the NoC router, we used ORION 3.0 [13].

TABLE II: NoC PLATFORM PARAMETERS

| Number of nodes | 64 |
|---|---|
| Network topology | 8 × 8 2D-Mesh |
| Virtual channels per port | 4 |
| Flow control | Credits-based |
| Flit size | 128 bits |
| Operating frequency | 2 GHz |
| Operating voltage | 1 V |
| Process technology | 45 nm |

The parameters of the NoC platform considered in our experiments are shown in Table II. We consider an 8×8 mesh NoC topology. Each node in the NoC is comprised of a processing element (PE) and a router. Each router can have a maximum of 4 input and 4 output ports besides the interface with the PE. Each port has 4 VCs with 4 flit buffers each. The table for storing Q-values has 63 rows, one for each destination, and 2 columns for Q-values for the horizontal and vertical directions. For the proposed routing algorithm, we include a fourth column to store route information of previous packet flows. Noxim was updated to use accurate energy estimates from the synthesized design for reading and updating the table for each Q-routing implementation (Q-RASP and comparison frameworks).

For comparison, we consider three different implementations of Q-routing in NoCs from prior work [2], [3], [6], the local congestion-based adaptive routing policy DyAD [14], and the popular dimension order routing policy XY. QR [2] uses the local queue latency of the packet as the cost for a routing decision. It does not use additional updates to the policy. BiLCQ [3] uses the queue length of the downstream router as the cost, while using an additional update to the policy for every routing action. The additional update is applied to the packet flow in the reverse direction of the packet being routed. CrQ [6] also uses a cost from the neighbor node like BiLCQ, except it uses the latency of the packet in the input queue. It also uses an additional table to implement a confidence-based learning rate for updating Q-values. DyAD [14] is an adaptive routing algorithm which uses the current queue length of the neighbors to make routing decisions without using Q-learning. All prior Q-routing policies use a learning rate ($\alpha$) of 0.50 as recommended in those works, and discount factor ($\gamma$) of 1. Our proposed policy uses a higher learning rate of 0.70, and a lower discount factor of 0.90. We increase the learning rate for faster convergence of Q-values and decrease the discount factor to allow the policy to switch to alternate paths faster. The regional cost factor $\mu$ is set to 0.1. We evaluate these policies using the metrics of average network latency, total network energy consumption, and area.

### B. Performance Results

We use average NoC packet latency for evaluating the performance of Q-RASP. Packet latency is defined as the duration in cycles from the time the packet is injected into the NoC at the source node to the time it is ejected at the destination node. It does not include the time spent in the injection queue. Using various synthetic traffic patterns, we monitor packet latency with increasing injection rates. As shown in Fig. 4(a), Q-RASP outperforms all other routing policies for bit-reversal, butterfly, and transpose traffic due to its path and region contention aware cost design and Shared Path Experience update mechanism. For random uniform traffic, BiLCQ performs better than Q-RASP. This can be attributed to the reverse path updates used by BiLCQ. As random uniform traffic utilizes the reverse of each path with equal probability, the additional updates from BiLCQ ensure that about half the paths have updated Q-values without any packets traversing those paths. Generally, Q-routing policies outperform XY and DyAD.

Fig. 4(b) shows the normalized average packet latency for applications from the PARSEC benchmark suite. Q-RASP achieves on average 18.3%, 15.7%, and 13.3% better packet latency than CrQ, QR, and BiLCQ respectively. In contrast to synthetic traffic, the gap between XY, DyAD, and Q-routing policies is smaller. In fact, for *bla*, XY performs the best after Q-RASP. The drop in Q-routing performance can be explained by the fact that real application traffic does not exhibit the permanent patterns found in synthetic traffic which can be learned by Q-routing methods. Q-RASP can overcome this by using experience sharing to accelerate learning for short-term patterns using additional updates.

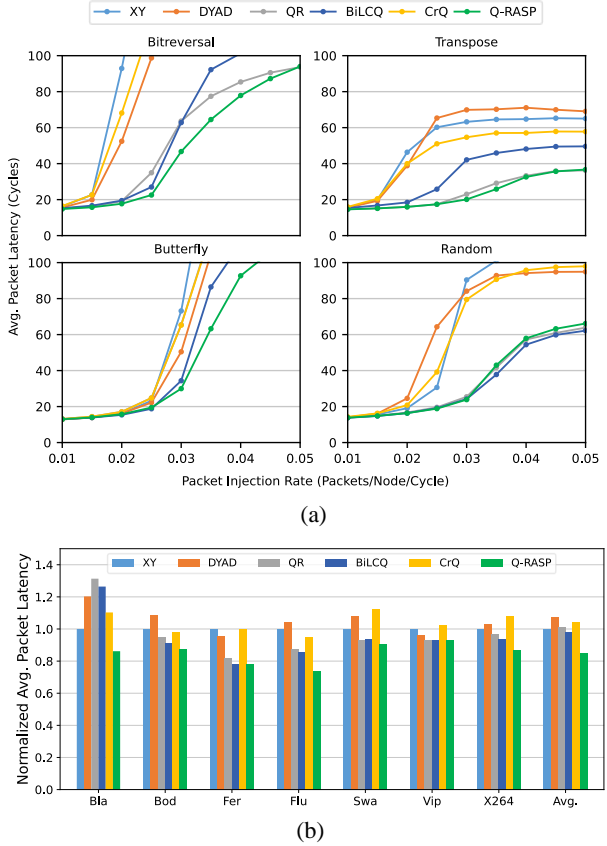

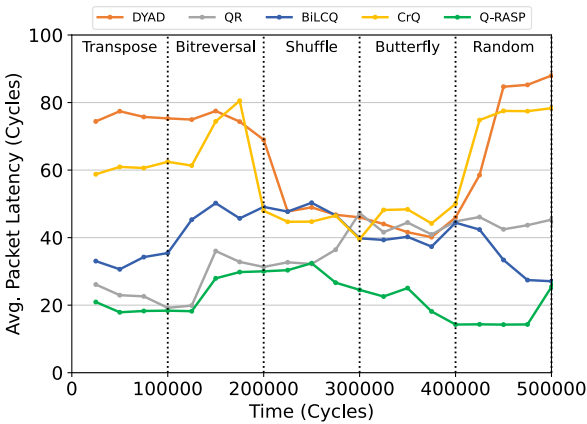

Figure 4. (a) Average packet latency for synthetic traffic patterns. (b) Normalized average packet latency for PARSEC applications.

Figure 5. Average network latency for interval-wise synthetic traffic.

While patterns in synthetic traffic are fixed, routing policies must be able to readjust to changing traffic patterns over time. To test the adaptability under varying traffic patterns, we set the traffic pattern to change after an interval of 100,000 cycles and monitored packet latency. The results for average packet latency shown in Fig. 5 demonstrate that Q-RASP can sustain low packet latency even under varying traffic pattern scenarios.

## C. Energy Results

Fig. 6(a) and Fig. 6(b) show the total NoC energy consumed per packet for synthetic traffic and PARSEC applications respectively, across all routing techniques. Q-RASP reduces total energy consumption per packet by 6.7%, 3.1%, and 2.3% compared to CrQ, QR, and BiLCQ respectively. It is able to reduce total energy consumption by 1) using a low-power Q-routing cost calculation method (Q-RASP cost calculation does not involve keeping track of queuing times of packets), 2) allowing lower precision Q-values by using a cost with a lower upper limit compared to packet latency, and 3) reducing packet latency using better routing decisions.

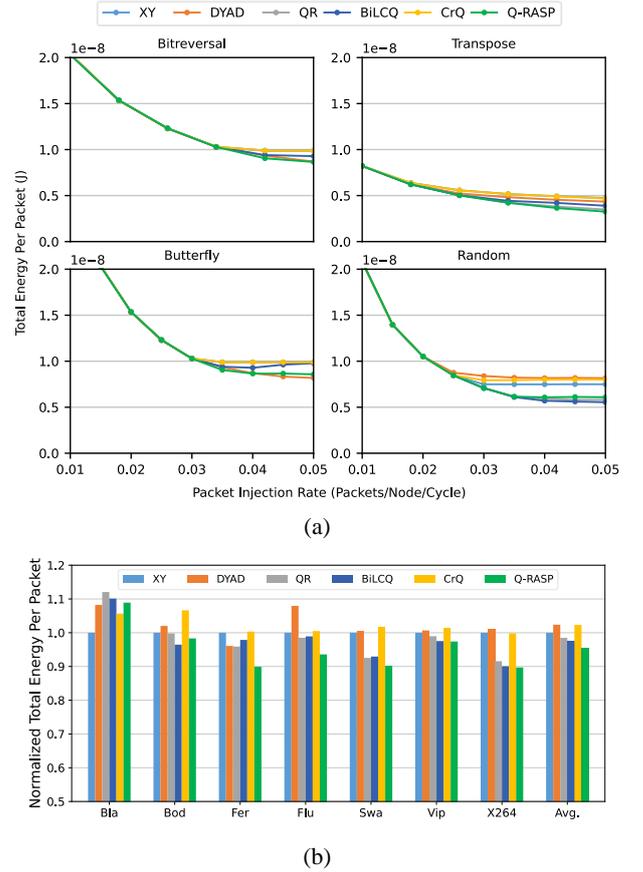

Figure 6. (a) Total energy per packet for synthetic traffic patterns. (b) Normalized total energy per packet for PARSEC applications.

## D. Area Results

Lastly, we present the area overhead of each Q-routing design in Table III. We minimize the area overhead in Q-RASP by 1) using lower precision Q-values, and 2) reducing the computational overhead of computing the cost by using contention-based metrics instead of tracking packet latency.

TABLE III. AREA OVERHEAD ANALYSIS FOR DIFFERENT Q-ROUTING IMPLEMENTATIONS

| Routing | Area Overhead ($\mu m^2$) | | | | |
|---|---|---|---|---|---|
| | *Router* | *Q Table* | *%* | *Cost Logic* | *%* |
| QR | 0.0426 | 0.0021 | 5.07% | 0.00026 | 0.62% |
| BiLCQ | 0.0426 | 0.0046 | 10.98% | 0.00001 | 0.03% |
| CrQ | 0.0426 | 0.0046 | 10.97% | 0.00026 | 0.62% |
| Q-RASP | 0.0426 | 0.0024 | 5.67% | 0.00001 | 0.02% |

## VI. CONCLUSION

In this work, we identify shortcomings in the design of existing Q-routing implementations in NoCs, arising due to suboptimal cost definition and inadequate policy updates. We present our novel Q-routing implementation (Q-RASP) to address these shortcomings by defining a path- and region-aware cost function. We also introduce a new update mechanism called Shared Path Experience to share the learning experience of packet flows using the same route for different destinations. Using Q-RASP, we improve average NoC packet latency by up to 18.3% and reduce total energy per packet by up to 6.7% with minimal area overhead compared to prior Q-routing implementations, highlighting the promise of our approach.

REFERENCES

[1] S. Pasricha, and N. Dutt, "On-Chip Communication Architectures", Morgan Kauffman, ISBN 978-0-12-373892-9, Apr 2008.
[2] J. Boyan and M. Littman, "Packet Routing in Dynamically Changing Networks: A Reinforcement Learning Approach," in *Advances in Neural Information Processing Systems*, Morgan-Kaufmann, 1993.
[3] F. Farahnakian *et al.*, "Bi-LCQ: A low-weight clustering-based Q-learning approach for NoCs," *Microprocess. Microsyst.*, vol. 38, no. 1, pp. 64–75, Feb. 2014.
[4] Y. Gupta and L. Bhargava, "Reinforcement Learning based Routing for Cognitive Network on Chip," in *Proc. 2nd Inter. Conf. on ICTCS*, New York, USA, Mar. 4-5, 2016.
[5] F. Farahnakian *et al.*, "Q-learning based congestion-aware routing algorithm for on-chip network," in *IEEE 2nd Inter. Conf. on Networked Embedded Systems for Enterprise Applications*, Perth, Australia, Dec. 8-9, 2011.
[6] N. Gupta *et al.*, "Improved Route Selection Approaches using Q-learning framework for 2D NoCs," in *Proc. of the 3rd Inter. Workshop on Many-core Embedded Systems*, New York, USA, Jun. 13-14, 2015.
[7] S. Srivastava *et al.*, "Intelligent congestion control for NoC architecture in Gem5 simulator," in *IEEE 15th Inter. Symp. on Embedded Multicore/Many-core Systems-on-Chip*, Penang, Malaysia, Dec. 19-22, 2022.
[8] H. Liu *et al.*, "TTQR: A Traffic- and Thermal-Aware Q-Routing for 3D Network-on-Chip," *Sensors*, vol. 22, no. 22, Jan. 2022.
[9] Y. Liu *et al.*, "A Q-Learning-Based Fault-Tolerant and Congestion-Aware Adaptive Routing Algorithm for Networks-on-Chip," *IEEE Embed. Syst. Lett.*, vol. 14, no. 4, pp. 203–206, Dec. 2022.
[10] E. Taheri, S. Pasricha, and M. Nikdast, "DeFT: A Deadlock-Free and Fault-Tolerant Routing Algorithm for 2.5D Chiplet Networks," in *DATE*, Antwerp, Belgium, Mar. 14-23, 2022.
[11] S. Pasricha *et al.*, "OE+IOE: a novel turn model based fault tolerant routing scheme for networks-on-chip," in *Proc. 8th IEEE/ACM/IFIP Inter. Conf. on Hardware/Software Codesign and System Synthesis*, Scottsdale, USA, Oct. 24-29, 2010.
[12] V. Catania *et al.*, "Cycle-Accurate Network on Chip Simulation with Noxim," *ACM Trans. Model. Comput. Simul.*, vol. 27, no. 1, p. 4:1-4:25, New York, USA, Aug. 2016.
[13] A. B. Kahng, B. Lin, and S. Nath, "ORION3.0: A Comprehensive NoC Router Estimation Tool," *IEEE Embed. Syst. Lett.*, vol. 7, no. 2, pp. 41–45, Jun. 2015.
[14] J. Hu and R. Marculescu, "DyAD: smart routing for networks-on-chip," in *Proc. 41st Annu. Design Automation Conf. (DAC)*, New York, USA, Jun. 7-11, 2004.